\documentclass[aps,pra,twocolumn,amsfonts,amssymb,amsmath,showpacs,
floatfix,nofootinbib,groupedaddress,superscriptaddress,citesort]{revtex4}
\usepackage{mathrsfs}
\usepackage{amsfonts}
\usepackage{amstext}
\usepackage{amsmath}
\usepackage{amssymb}\usepackage{subfigure}
\usepackage{bm}
\usepackage{bbm}\usepackage{threeparttable}

\usepackage{graphicx}
\def\qed{\leavevmode\unskip\penalty9999 \hbox{}\nobreak\hfill
     \quad\hbox{\leavevmode  \hbox to.77778em{%
              \hfil\vrule   \vbox to.675em%
               {\hrule width.6em\vfil\hrule}\vrule\hfil}}
     \par\vskip3pt}

\begin{document}

\title{Features of preparable entangled  states in Gaussian quantum networks}

\author{Shuanping Du}
 \affiliation{School of Mathematical
Sciences, Xiamen University, Xiamen, Fujian, 361000, China}

\author{Zhaofang Bai}\thanks{Corresponding author}
\email{baizhaofang@xmu.edu.cn} \affiliation{School of Mathematical
Sciences, Xiamen University, Xiamen, Fujian, 361000, China}

\begin{abstract}
Large-scale quantum networks have been employed to overcome practical constraints on transmission and
storage for single entangled systems.
The deterministic preparation of entangled states is one of the key factors for realization of quantum networks.
There is no efficient method to verify whether single multipartite entanglement can be prepared by multisource quantum networks.
Here, we theoretically analysize under what conditions entangled states can be prepared in three kinds of basic Gaussian quantum networks, named  triangle networks, star-shaped networks and chain-type networks. Some necessity criteria are derived for all preparable  entangled Gaussian states in such networks. It shows that the network structure imposes strong constraints on the set of preparable entangled Gaussian states, which is fundamentally different with the standard single multipartite entanglement. This takes the first step towards
understanding network mechanism for preparing entangled Gaussian states.

\end{abstract}
\pacs{03.65.Ud, 03.67.-a, 03.67.Mn.}
\maketitle

\maketitle

${Introduction}.$ Quantum networks feature independent sources distributing
particles to different parties in the network, which may have the huge potential to
to enhance the performance of various quantum technologies by distributing entangled states over long distances
 in the near future \cite{Kimble,Komar,Guerin,Wehner,Tavakoli}.
 Despite facing many technological challenges,  basic quantum networks have been implemented and some  progress have been made on some physical platforms where matter and light interact \cite{Cirac,Duan,Tanzilli,Chaneliere,Pan1,Sasaki,Boaron,Liw,Pan2}.

In the spirit of entanglement swapping, the entanglement initially prepared on links of quantum networks can then be propagated to  entire networks by fulfilling entangled measures at the particles \cite{Acin,Perseguers1,Perseguers2}. This results in very strong forms of multipartite quantum correlations spreading across the entire quantum networks. Characterizing such quantum correlations is a natural question with clear basic interest. It impacts technological developments of future quantum networks and also helps researchers in understanding  multipartite quantum correlations in more sophisticated and qualitative scenarios.

The first step has been taken to characterize multipartite quantum correlations in discrete-variable quantum networks. The
Bell locality \cite{Bell} has been generalized to quantum networks \cite{Branciard,Fritz,Rosset}. The vital idea lies in that different sources in the network
distributing physical systems to the parties should be assumed to be independent. This is the departure of standard Bell nonlocality.
As a typical feature, two independent parties in a chain-shaped quantum network consisting of two entangled systems can generate a new entanglement by
local operations and classical communication \cite{Zukowski,Skrzypczyk}. Although major advances have been seen in
characterizing multipartite quantum correlations of quantum networks for the last few years \cite{Henson,Chaves,luo1,luo2,Wolfe,Renou,Pozas,Kraft,Luo3,Xu,Weinbrenner,Wang}, as far as our understanding of multipartite quantum correlations in networks is very limited, characterizing multipartite quantum correlations is still in the early stages.

The existing achievements on multipartite quantum correlations
focus on discrete-variable systems, which are natural
for computing. However, light as the only carrier of quantum
information in network transmission, is bosonic and requires a
continuous-variable system.
This motivates us to study multipartite quantum correlations in Gaussian quantum networks. It is known that Gaussian quantum networks serve as the basis for continuous-variable quantum information processing \cite{Jouguet,Ghalaii,Liu,Motaharifar,Bae,Pan3}.
The paper is to study multipartite quantum  correlations from the point of entanglement.
Inspired by recent development of quantum network entanglement theory \cite{Luo3,Kraft2,Navascues}, we investigate  preparation of multipartite entanglement in Gaussian quantum networks, that is, which entangled states can be prepared in Gaussian quantum networks. Our goal is to reveal the possibility and constraint conditions for entanglement preparation under triangle networks, star-shaped networks and chain-type networks (see Fig.1).
Our study provides a complement to discrete-variable systems counterparts and also provide insights into continuous-variable Gaussian quantum networks. This takes the first step towards understanding the mechanism for preparing entangled Gaussian states in Gaussian quantum networks.

\vspace{0.1in}

${Gaussian\  states\  and\ Gaussian\ unitary\ operations}.$
Gaussian  states and  Gaussian unitary operations are primary tools for dealing with continuous-variable quantum information processing \cite{Cerf,Serafini}. Gaussian states constitutes versatile resources for quantum communication with bosnic continuous-variable systems and  are easy to prepare and control in a range of setups including primarily quantum optics, atomic ensembles, trapped ions, opto-mechanics, as well as networks interfacing these diverse technologies.
Gaussian unitary operations can be realized as a passive operation, a single-mode squeezing operation
on each of the $m$ modes, and a subsequent second passive operation.
Here, we provide exact definitions on bosonic Gaussian states and Gaussian unitary operations (see \cite{Weedbrook} for a review).
Let ${\mathcal H}$ be
an infinite dimensional Hilbert space with the fixed Fock basis $\{|n\rangle\}_{n=0}^{+\infty}$.
When we consider the $m$-mode continuous-variable systems ${\mathcal H}^{\otimes m}$, we adopt $(\{|n\rangle\}_{n=0}^{+\infty})^{\otimes m}$
as its reference basis.
For a quantum state $\rho\in {\mathcal H}^{\otimes m}$, the characteristic function of $\rho$ is defined as
$$\begin{array}{lll}
{\mathcal X}_\rho(\lambda) & = & tr(\rho D(\lambda)),\\
D(\lambda) & = & \otimes _{i=1}^m D(\lambda_i),\\
D(\lambda_i) & = & e^{(\lambda_i\widehat{a_i}^\dag-{\overline \lambda_i}\widehat{a_i})},
\end{array}\eqno (1)$$
here $\widehat{a_i}$ and $\widehat{a_i}^\dag$ are the annihilation operator and the creation operator in mode $i$, \quad $\lambda=(\lambda_1, \cdots, \lambda_m)^t, \quad  {\overline \lambda_i}$ denotes the complex conjugate of $\lambda_i$. Gaussian states are those states for which ${\mathcal X}_\rho(\lambda)$  is a Gaussian function of the phase space, i.e.,
$${\mathcal X}_\rho(\lambda)=exp^{-\frac{1}{4}{\overrightarrow r}\Omega V\Omega^t{\overrightarrow r}^t-i(\Omega d)^t{\overrightarrow r} ^t},\eqno (2)$$ where $\overrightarrow{r}=(\lambda_{1x}, \lambda_{1y}, \ldots,\lambda_{mx}, \lambda_{my} )$, $\lambda_{jx},\lambda_{jy}$ are the real part and imaginary part of $\lambda_j \ (j=1,2, \ldots, m)$, V is a $2m\times 2m$ real hermitian matrix which is called covariance matrix satisfying the uncertainty relation $V+i\Omega\geq 0$, $d\in {\mathbb R}^{2m}$ is called mean value,
$\Omega=\bigoplus_{k=1}^m \left(\begin{array}{cc}
                                  0 & 1\\
                                  -1 & 0\end{array}\right)$. Note that $\det V \geq 1$ and $\det V = 1$ if and only if $\rho$ is pure. It is clear that $(V,d)$
can describe Gaussian state $\rho$ completely. So $\rho$ can be usually written in $\rho(V, d)$. As a shift of the mean value amounts to a local unitary operation that leaves entanglement invariant, we will assume our Gaussian states to have zero mean value, $d=0 $, without loss of generality.

Gaussian unitary operations are unitary operations  that transform Gaussian states into Gaussian states. Explicitly, $$U=\exp^{-i\hat{H}}, \eqno (3)$$ with a generic quadratic Hamiltonian  $$\hat{H}=\hat{\xi}^\dag H {\hat{\xi}}, \eqno (4)$$
$\hat{\xi}=(\widehat{a_1}, \cdots, \widehat{a_m}, {\widehat{a_1}}^\dag, \cdots,{\widehat{a_m}}^\dag)^\dag,$ $H$ is complex $2m\times 2m$ hermitan matrix.

\vspace{0.1in}

${Gaussian\  quantum\  networks}\ (GQNs).$
The quantum network is Gaussian if the operations associated with links at the parties are Gaussian unitary, and the final state shared by entire network is Gaussian \cite{Ghalaii}. There are three kinds of basic GQNs named triangle networks, star-shaped networks and chain-type networks.
The triangle network has three parties $A, B, C$ which are connected pairwise by three two-mode entangled Gaussian source states, so the network forms a triangle and all parties receive two infinite dimensional quantum systems.  We apply a local Gaussian unitary operation on every party which results in a global entangled state on the entire network (see Fig.1(a)). The star-shaped network has $n+1$ parties $A_1, A_2, \cdots, A_{n+1}$ and
 a central party ( referred to as $A_1$) shares a two-mode entangled Gaussian state with each of the $n$ parties, i.e., Gaussian states are provided by $n$ independent sources, hence the network forms a star. Each source produces a two-mode Gaussian state which acts on the separable infinite dimensional complex Hilbert space:  each entangled states $\rho_i$ is shared by $A_1$ and $A_i\  (2\leq i\leq n+1)$. Thus $A_1$ receives $n$ infinite dimensional quantum systems and each $A_i\ (i\geq 2 )$ receives one infinite dimensional quantum system.
 Finally, each party can apply a local Gaussian unitary operation to their systems, which we denote with $U_{A_1}, U_{A_2}, \cdots, U_{A_{n+1}}$. This results in a global state for the entire Gaussian network (see Fig.1(b)). The chain-type quantum network contains $n+1$ parties, denoted by $A_1, A_2, \cdots, A_{n+1}$, where adjacent parties $A_{i}$
and $A_{i+1}$ share a two-mode entangled Gaussian state $\rho_i \ (1\leq i\leq n)$, and the net forms a chain. Therefore both $A_1$ and $A_{n+1}$ receive one infinite dimensional quantum system, each $A_i\ (2\leq i\leq n )$ receives two infinite dimensional quantum systems, and each party  applies a local Gaussian unitary operation to their systems, which we also denote with $U_{A_1}, U_{A_2}, \cdots, U_{A_{n+1}}$. This outputs  a global state for the entire Gaussian network (see Fig. 1(c)).

The central question we consider is which entangled Gaussian  state admits a decomposition of the form
$$(U_{A_1}\otimes\cdots\otimes U_{A_{n+1}})(\rho_1\otimes\cdots \otimes\rho_s)(U_{A_1}^\dag\otimes\cdots\otimes U_{A_{n+1}}^\dag),\eqno(5)$$ each $\rho_i (1\leq i\leq s)$ is a shared 2-mode entangled Gaussian states and $U_{A_i} (1\leq i\leq n)$ is a local Gaussian unitary operation on the party $A_i$.
Ours aim is to reveal under what conditions multipartite entangled Gaussian states can be prepared by source states with entanglement depth 2 \cite{Navascues}.
For clarity, let ${\mathcal {GQN}}_{\triangle}$, ${\mathcal {GQN}}_{\star}$ and ${\mathcal {GQN}}_{\S}$ denote the set of  preparable entangled Gaussian states
on the triangle network, the star-type network and the chain-type network, respectively.


\vspace{0.1in}



\begin{figure}[htbp]
	\centering\vspace{-0.35cm}\subfigbottomskip=1pt
\subfigure{	\label{(a)}\includegraphics[scale=0.35]{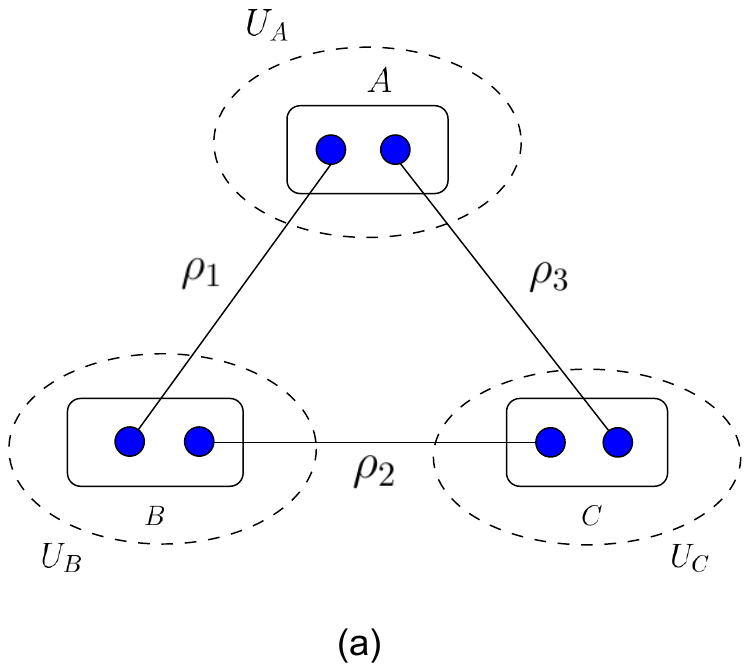}}
\vspace{-5.0cm}

\subfigure{\includegraphics[scale=0.35]{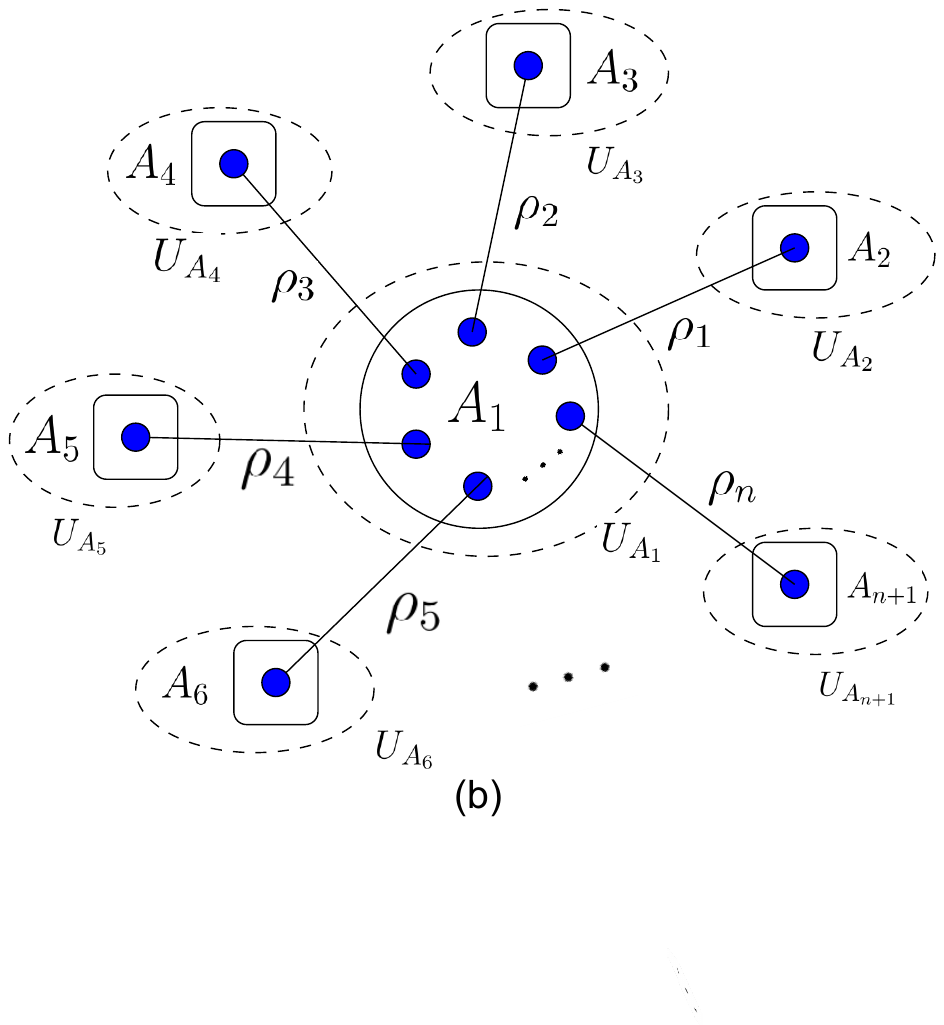}}

\vspace{-5.4cm}\subfigure{	\includegraphics[scale=0.35]{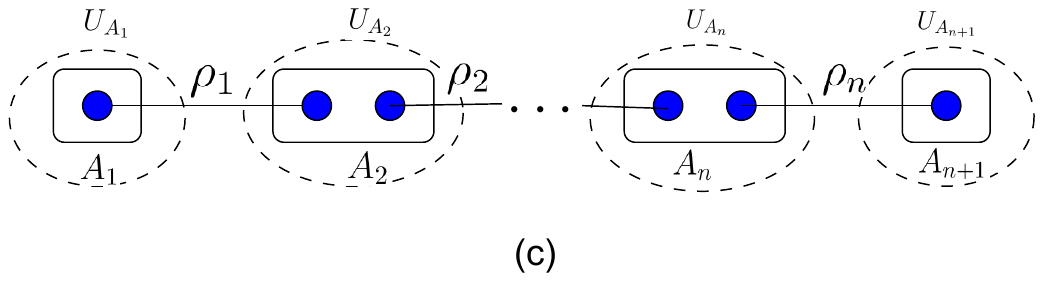}}

	\vspace{-7.35cm}\caption{\small Diagram for preparing entangled states in Gaussian quantum networks. Each $\rho_i$ is a $2$-mode entangled Gaussian state shared by two parties, $U_{A_i}$ is a local Gaussian unitary operation on the party $A_i$, a blue ball denotes an infinite dimensional quantum system.}
\end{figure}

${Results}.$ We discuss  necessity criteria of entanglement preparation in terms of the mutual information, the universal extension of the squashed entanglement and a computable Gaussian quantum correlation measure.
we need firstly to extend the concept of bipartite mutual information to multipartite mutual information. This is particular important to study network quantum correlation which is always multipartite.
Recall that for a bipartite Gaussian state $\rho_{AB}$, the mutual information between subsystems $A$ and $B$ is defined as \cite{Adesso3} $$\begin{array}{lll}
&{\mathcal I}_2(A:B)\\=&{\mathcal S}_2(\rho_A)+{\mathcal S}_2(\rho_B)-{\mathcal S}_2(\rho_{AB})\\
                    =& \frac{1}{2}[\ln(\det V_A)+\ln(\det V_B)-\ln(\det V)],\end{array} \eqno (6)$$ where ${\mathcal S}_2(\rho_{AB})=-\ln\rm{tr}(\rho_{AB}^2)=\frac{1}{2}\ln(\det V)$ is the Renyi-$2$ entropy,
                    $V_A$, $V_B$ and $V$ denotes the covariance matrix of $\rho_A$, $\rho_B$ and $\rho_{AB}$, respectively. Following the spirit from Cerf and Adami \cite{Cerf2}, for a tripartite Gaussian state $\rho_{ABC}$, we define tripartite mutual information as $$\begin{array}{ll}
                  &  {\mathcal I}(A:B:C)\\=&{\mathcal I}_2(A:B)-{\mathcal I}_2(A:B|C)\\
=&{\mathcal S}_2(\rho_A)+{\mathcal S}_2(\rho_B)+{\mathcal S}_2(\rho_C)\\
&-{\mathcal S}_2(\rho_{AC})-{\mathcal S}_2(\rho_{AB})-{\mathcal S}_2(\rho_{BC})\\
&+{\mathcal S}_2(\rho_{ABC})\\
=&\frac{1}{2}[\ln(\det V_A)+\ln(\det V_B)+\ln(\det V_C)\\
&-\ln(\det V_{AC})]-\ln(\det V_{AB})-\ln(\det V_{BC})\\
&+\ln(\det V)]
,\end{array} \eqno (7) $$ here ${\mathcal I}_2(A:B|C)={\mathcal S}_2(\rho_{AC})+{\mathcal S}_2(\rho_{BC})-{\mathcal S}_2(\rho_{ABC})-{\mathcal S}_2(\rho_C)$ is the conditional mutual information of $\rho_{ABC}$,  $V_A$, $V_B$, $V_C$, $V_{AB}$, $V_{AC}$,$V_{BC}$ and $V$ denotes the covariance matrix of $\rho_A$, $\rho_B$, $\rho_C$, $\rho_{AB}$, $\rho_{AC}$, $\rho_{BC}$ and $\rho_{ABC}$, respectively.
 More general,
for a multipartite  Gaussian state $\rho_{A_1, A_2, \cdots, A_n}$, we define the mutual information of $\rho_{A_1, A_2, \cdots, A_n}$ as follows:
$$\begin{array}{ll}
&{\mathcal I}(A_1:A_2:\cdots :A_n)\\=&{\mathcal S}_2(\rho_{A_1})+\cdots +{\mathcal S}_2(\rho_{A_n})\\
                                   &-{\mathcal S}_2(\rho_{A_1A_2})-\cdots-{\mathcal S}_2(\rho_{A_1A_n})\\
                                   &-{\mathcal S}_2(\rho_{A_2A_3})-\cdots-{\mathcal S}_2(\rho_{A_2A_n})\\
                                   & - \cdots -{\mathcal S}_2(\rho_{A_{n-1}A_n})\\
                                   &+\sum_{1\leq i\neq j\neq k\leq n} {\mathcal S}_2(\rho_{A_iA_jA_k})\\
                                   & -\cdots +(-1)^{n-1}{\mathcal S}_2(\rho_{A_1 A_2 \cdots A_n}).\end{array}\eqno (8)$$
We remark that ${\mathcal I}(A_1:A_2:\cdots :A_n)$ is determined by the covariance matrix of $\rho_{A_1, A_2, \cdots, A_n}$. It can be expressed in terms of determinants of the local covariance matrices for the reduced state of $\rho_{A_1, A_2, \cdots, A_n}$.
From the unitary invariance of the Renyi-$2$ entropy and the Schmidt decomposition of continuous-variable systems \cite{Blanchard},
${\mathcal I}(A_1:A_2:\cdots :A_n)$  bears the following properties:

(i) ${\mathcal I}(A_1:A_2:\cdots :A_n)$ is invariant under local unitary operations.

(ii) ${\mathcal I}(A_1, A_2, \cdots, A_n)= 0$ for any pure Gaussian state in an odd partite system.




\vspace{0.1in}
From Fig. 1(a), Fig. 1(b) and Fig. 1(c), the intuition is that the amount of global correlations for any  $\rho\in{\mathcal GQN}_\star$ ( $\rho\in{\mathcal GQN}_\S$ or $\rho\in{\mathcal GQN}_\triangle$)
ought to be limited since all parties do not share any common multipartite information. This intuition can be made formal by considering multipartite mutual information for Gaussian networks states.
\vspace{0.1in}

\textbf{Theorem 1.} {\it For any preparable entangled Gaussian state $\rho$, the following statements hold true:
$$(i)\  If\  \rho\in{\mathcal {GQN}}_\triangle, \mbox{we have}\  {\mathcal I}(A:B:C)=0.\hspace{0.6in}$$}
$$(ii)\ If\  \rho\in{\mathcal {GQN}}_\star,
we \ have \ {\mathcal I}(A_1:A_2:\cdots :A_{n+1})=0.$$
$$(iii)\ If\  \rho\in{\mathcal {GQN}}_\S, we \ have \ {\mathcal I}(A_1:A_2:\cdots :A_{n+1})=0.$$

{\bf Proof.}  We only treat the star-shaped network and the other two scenarios can be proved similarly.  By the property (i) of multipartite mutual information, we can assume $\rho=\rho_1\otimes \rho_2\otimes \cdots \otimes \rho_n$, one can check $$\begin{array}{ll}
\rho_{A_1}&=\rm{tr} _{A_2}\rho_1\otimes \rm{tr} _{A_3}\rho_2\otimes\cdots \otimes \rm{tr} _{A_{n+1}}\rho_n,\\
\rho_{A_i}&=\rm{tr}_{A_1}\rho_{{i-1}}\ (2\leq i\leq n+1),\\
\rho_{A_1A_2}&=\rho_1\otimes \rm{tr}_{A_3}\rho_2\otimes \cdots \otimes \rm{tr}_{A_{n+1}}\rho_n,\\
&\ \vdots\\
\rho_{A_1A_n}&=\rm{tr}_{A_2}\rho_1\otimes \rm{tr}_{A_3}\rho_2\otimes \cdots \otimes \rho_n,\\
\rho_{A_iA_j}&=\rm{tr}_{A_1}\rho_{i-1}\otimes \rm{tr}_{A_1}\rho_{j-1}\ (i\neq j\geq 2),\\
\rho_{A_1A_2A_3}&=\rho_1\otimes \rho_2\otimes \rm{tr}_{A_4}\rho_3\otimes \cdots \otimes \rm{tr}_{A_{n+1}}\rho_n,\\
&\ \vdots\\
\rho_{A_1A_nA_{n+1}}&=\rm{tr}_{A_2}\rho_1\otimes \rm{tr}_{A_3}\rho_2\otimes \cdots \otimes\rho_{n-1}\otimes \rho_n,\\
\rho_{A_iA_jA_k}&=tr_{A_1}\rho_{i-1}\otimes \rm{tr}_{A_1}\rho_{j-1}\otimes \rm{tr}_{A_1}\rho_{k-1}\\
 & \ \ \ \ \ \ \ \ \ \ \ \hspace{0.1in}\ \ \  (2\leq i\neq j\neq k\leq n+1)\\
 &\ \vdots\\
 \rho_{A_1A_2\cdots A_n} &=\rho_1\otimes \rho_2\otimes \cdots \otimes \rho_n.\\
\end{array}
\eqno(9)$$
Let $V_{i(i+1)}$ be the covariance matrix of $\rm{tr}_{A_{i+1}}\rho_i$, $V_{i1}$ be the covariance matrix of $\rm{tr}_{A_1}\rho_{i-1} (2\leq i\leq n)$, and $V_{i}$ be
be the covariance matrix of $\rho_i$, one can see that ${\mathcal I}(A_1:A_2:\cdots :A_{n+1})$ is determined by  $\ln(\det V_{i(i+1)})$, $\ln(\det V_{i1})$, and $\ln(\det V_{i})$ from (8). In the following, we show that every coefficient of the three determinants actually equals zero and so
 ${\mathcal I}(A_1:A_2:\cdots :A_{n+1})=0$. By our assumption,
 the covariance matrix of Gaussian state $\rho_{A_1}$ is
$\oplus_{i=1}^nV_{i(i+1)}$. Therefore $${\mathcal S}_2(\rho_{A_1})=\frac{1}{2}[\ln(\det V_{12})+\cdots+\ln(\det V_{n(n+1))}]. \eqno (10)$$
We now focus on coefficients of $\ln(\det V_{12})$ in ${\mathcal I}(A_1:A_2:\cdots :A_{n+1})$. A direct computation on the bipartite Renyi-2 entropy shows that only $${\mathcal S}_2(\rho_{A_1A_3}), {\mathcal S}_2(\rho_{A_1A_4}), \cdots, {\mathcal S}_2(\rho_{A_1A_{n+1}})\eqno (11)$$ contain $\ln(\det V_{12})$, this implies the coefficient of  $\ln(\det V_{12})$ for the bipartite Renyi-2 entropy is $$-C_{n-1}^1=-(n-1).\eqno (12)$$ Similarly, by computing the tripartite Renyi-2 entropy, only $${\mathcal S}_2(\rho_{A_1A_iA_j})(3\leq i\neq j\leq n+1)\eqno (13)$$ contain the term $\ln\det V_{12}$ and so the coefficient of $\ln\det V_{12}$ is $$C_{n-1}^2=\frac{1}{2}(n-1)(n-2).\eqno (14)$$ Using analogous treatments, we can obtain the coefficient of $\ln\det V_{12}$ in ${\mathcal I}(A_1:A_2:\cdots :A_{n+1})$ is $$C_{n-1}^0-C_{n-1}^1+C_{n-1}^2-\cdots +(-1)^{n-1}C_{n-1}^{n-1}.\eqno (15)$$ Note that the formula (15) actually equals to zero, hence the term
$\ln\det V_{12}$ in ${\mathcal I}(A_1:A_2:\cdots :A_{n+1})$ disappears. Similarly, one can see all other terms also equal to zero and so
${\mathcal I}(A_1:A_2:\cdots :A_{n+1})=0$.

\vspace{0.1in}

Theorem 1  shows that the multipartite mutual information being zero is necessary for
entangled Gaussian states which can be prepared by GQNs.  The constraint condition
shows the network structure imposes strong constraints on the set of preparable entangled Gaussian states. This
contributes to a deeper understanding of the fundamental network mechanism for preparable entangled Gaussian states, which is fundamentally different with the standard single multipartite entanglement.

In the realm of continuous-variable quantum information processing, fully symmetric  Gaussian states
are the privileged resources for most communication protocols \cite{Renner} and are also basic test grounds for investigating
structural aspects of multipartite continuous-variable entanglement \cite{Adesso5,Adesso6,Loock}. A fundamental question is that whether fully symmetric  Gaussian states can be prepared by GQNs. Theorem 1  provides a tool to verify whether single multipartite entanglement can be prepared by multisource quantum networks.
For any $n+1$-partite entangled Gaussian state $\rho$, if ${\mathcal I}(A_1:A_2:\cdots :A_{n+1})\neq 0$, then $\rho$ can not be prepared.
Recall that the symmetric spectrum of fully symmetric $n$-mode Gaussian state has the following structure: $$\begin{array}{ll}
v_-^2&=(b-e_1)(b-e_2),\\
v^2_{+^{(n)}}&=[b+(n-1)e_1][b+(n-1)e_2],\end{array}\eqno(16)$$  here the $v_-$ is the $n-1$-degenerate eigenvalue \cite{Adesso5}.
Let $|\psi\rangle$ be a $6$-mode and $4$-partite pure symmetric Gaussian state, the covariance matrix of $|\psi\rangle$
has the form $$V=\left(\begin{array}{cccc}
                       \sigma &\epsilon & \cdots & \epsilon\\
                       \epsilon & \sigma & \epsilon &\vdots\\
                       \vdots & \epsilon & \ddots & \epsilon\\
                       \epsilon & \cdots & \epsilon & \sigma\end{array}\right)_{6\times 6}, \eqno (17)$$ where $\sigma$ and $\epsilon$ are $2\times 2$ diagonal matrices, $\sigma=\text {diag}(b,b), \epsilon=\text {diag}(e_1,e_2), b\geq 1$.
It is evident that $$\begin{array}{ll}
v_-^2&=(b-e_1)(b-e_2),\\
v^2_{+^{(6)}}&=[b+5e_1][b+5e_2],\end{array}\eqno (18)$$ are the symplectic spectrum of $V$ which consists of $v_+$ and $5$ degenerate eigenvalue $v_-$. Imposing  the constraint of pure state $(v_-^2=v_{+^{(6)}}^2=1)$, one can further get
$$(e_1+e_2)b=-4e_1e_2.\eqno(19)$$
$$e_i=\frac{4b^2-4-(-1)^i\sqrt{(b^2-1)(36b^2-16)}}{10b}.\eqno(20)$$
A direct computation shows that
 $${\mathcal I}(A_1:A_2:A_3:A_4)= 0\Leftrightarrow b^6v_-^4v^2_{+^{(3)}}=v_-^{18}v^6_{+^{(4)}}.\eqno(21)$$
Combining (19) with (21), one can obtain $${\mathcal I}(A_1:A_2:A_3:A_4)= 0\Leftrightarrow b^6(b^2-4e_1e_2)=(b^2-3e_1e_2)^3.\eqno(22)$$
By Eq. (20), we have $$e_1e_2=\frac{1-b^2}{5}.\eqno(23)$$
Substituting $(23)$ into $(22)$, we can deduce $$\begin{array}{ll}{\mathcal I}(A_1:A_2:A_3:A_4)= 0\Leftrightarrow &225b^8-612b^6+576b^4\\
&-216b^2+27=0.\end{array}\eqno(24) $$
Let $b^2=t(t\geq 1)$, then $$f(t)=225t^4-612t^3+576t^2-216t+27=0.\eqno(25)$$ It is easy to check that $f'(1)=0$ and $f''(t)> 0$. Thus $f'(t)\geq 0 $ and so $f(t)$ is strictly increasing.
This implies that $$225b^8-612b^6+576b^4-216b^2+27=0\Leftrightarrow b=1.\eqno(26)$$
Using Eq. (20) again, we have $e_1=e_2=0$.  
Therefore $${\mathcal I}(A_1:A_2:A_3:A_4)= 0\Leftrightarrow b=1, e_1=e_2=0. \eqno(27)$$
This shows that if the mutual information of $|\psi\rangle$ equals $0$, then $|\psi\rangle$ is actual a $4$-partite product state.
The chain-type net case can be treated analogously.
By Theorem 1,  any $6$-mode and $4$-partite pure symmetric entangled Gaussian state can not be prepared by the star-type network
and the chain-type net. Fully
symmetric pure Gaussian states belong to the class of
continuous-variable Greenberger-Horne-Zeilinger (GHZ) type state \cite{Adesso5,Loock}, a parallel result in discrete-variable
systems is that all permutationally symmetric entangled
pure states can not be prepared under shared randomness
without classical communication \cite{Luo3}.
For general multimode multipartite mixed Gaussian states, there are finite number of possible values (depending the number of mode) such that the mutual
information of the states equals zero for fixed $e_1, e_2$. Thus almost all fully symmetric mixed entangled Gaussian states can not be prepared by GQNs.
\vspace{0.1in}



Beyond mutual information, combining Theorem 1 and  Fig. 1(a), we have an intuition  that entanglement on the bipartition $X|YZ$ is the sum of the entanglement of $X|Y$ and $X|Z$ in reduced states, where $X|YZ$ denotes all bipartions of $A|BC, B|AC$, and $C|AB$.
From Fig. 1(b), our intuition  is that entanglement on the bipartition $A_1|A_2\cdots A_{n+1}$ ought to be equal to the sum of the entanglement in the reduced states, i.e., $A_1|A_2, A_1|A_3, \cdots, A_1|A_{n+1}$. From Fig. 1(c), we also have an intuition that entanglement on the bipartition $A_i|A_1A_2\cdots A_{i-1}A_{i+1}\cdots A_n (2\leq i\leq n-1)$ should be equal to the sum of the entanglement $A_{i-1}|A_i$ and $A_{i}|A_{i+1}$ of reduced states.
We next make these formal by using universal extension of the squashed entanglement in continuous-variable systems

Recall that the squashed entanglement of a state $\rho_{AB}$ of a finite dimensional system $AB$ is defined as
$$E_{sq}(\rho_{AB})=\frac{1}{2}\inf_{\rho_{ABE}} {\mathcal I}(A:B|E), \eqno (28)$$ where infimum is over all extensions $\rho_{ABE}$ of the state $\rho_{AB}$ and ${\mathcal I}(A:B|E)$ is the conditional mutual information of $\rho_{ABE}$ \cite{Christandl}. The squashed entanglement is the only known entanglement measure satisfying all desired properties, such as monotonicity under selective unilocal operations, additivity for product states, monogamy and so on.
One can construct its extensions to the $2$-mode continuous-variable systems $AB$ as follows:
$$\hat{E}_{sq}(\rho_{AB})=\sup_{P_A, P_B} E(P_A\otimes P_B\rho_{AB}P_A\otimes P_B),\eqno (29)$$ where supremum is over all finite rank projectors $P_A$ and $P_B$  \cite{Shirokov}. It is known that $\hat{E}_{sq}(\cdot)$ is a nice entanglement measure of continuous-variable systems and also possess  all desire properties as $E_{sq}(\cdot)$.

 We remark that although the universal extension of the squashed entanglement in $2$-mode and bipartite systems has been provided \cite{Shirokov},
such extension is actually valid for multimode and bipartite systems by adjusting the choice
of finite rank projectors. Indeed, Proposition 5 of \cite{Shirokov} shows that $$\hat{E}_{sq}(\rho_{AB})=\lim_{n\rightarrow \infty}E(P_A^n\otimes P_B^n\rho_{AB}P_A^n\otimes P_B^n)\eqno(30)$$ for arbitrary sequence $\{P_A^n\}$ and $\{P_B^n\}$ of finite rank projectors strongly converging to the identity operators $I_A$ and $I_B$.
If both ${\mathcal H}_A$ and ${\mathcal H}_B$ are multimode systems, assume ${\mathcal H}_A={\mathcal H}_{A_1}\otimes \cdots\otimes {\mathcal H}_{A_s}$,
${\mathcal H}_B={\mathcal H}_{B_1}\otimes \cdots \otimes{\mathcal H}_{B_t}$, one can choose finite rank projectors $P_{A_i}^n (1\leq i\leq s)$ and $P_{B_j}^n (1\leq j\leq t)$ strongly converging to the identity operators $I_{A_i}$ and $I_{B_j}$.
It is easy to check that  $\otimes_{i=1}^sP_{A_i}^n$ and $\otimes_{j=1}^tP_{B_j}^n$ are finite rank projectors and strongly converges to the identity operator $I_A$ and $I_B$. Applying Proposition 5 of [55], we know the multimode generalization of the squashed entanglement is valid.


This point is paticularly important for studying the network entanglement which is always multimode. Combining the additivity for product states  of $\hat{E}_{sq}(\cdot)$ in multimode and bipartite systems \cite{Shirokov}, we have the following results.
\vspace{0.1in}

\textbf{Theorem 2.} {\it For any $\rho\in{\mathcal {GQN}}_\triangle$, we have $$\hat{E}_{sq}(\rho_{X|YZ}) =\hat{E}_{sq}(\rho_{XY})+ \hat{E}_{sq}(\rho_{XZ}),$$
here  $XYZ$ is any permutation of $ABC$,
$\hat{E}_{sq}(\rho_{XY})$ is the entanglement measure of a composite quantum system with respect to bipartite cut between $X$ and $Y$.}

\vspace{0.1in}

\textbf{Theorem 3.} {\it For any $\rho\in{\mathcal {GQN}}_\star$, we have $$\hat{E}_{sq}(\rho_{A_1|A_2\cdots A_{n+1}})=\hat{E}_{sq}(\rho_{A_1A_2})+\cdots +\hat{E}_{sq}(\rho_{A_1A_{n+1}}),$$  here  $\hat{E}_{sq}(\rho_{A_1A_{i}}) (2\leq i\leq n+1)$ is the entanglement measure of a composite quantum system with respect to bipartite cut between $A_1$ and $A_i$.}

\vspace{0.1in}

\textbf{Theorem 4.} {\it For any $\rho\in{\mathcal {GQN}}_\S$, we have $$\hat{E}_{sq}(\rho_{A_i|A_1\cdots A_{i-1}A_{i+1}\cdots A_{n+1}}) =\hat{E}_{sq}(\rho_{A_{i-1}A_i})+ \hat{E}_{sq}(\rho_{{A_i}A_{i+1}}),$$  here  $\hat{E}_{sq}(\rho_{A_{i-1}A_{i}}) (2< i< n+1)$ is the entanglement measure of a composite quantum system with respect to bipartite cut between $A_{i-1}$ and $A_i$.}

\vspace{0.1in}

Theorems 2, 3 and 4 show strong monogamy relations of Gaussian networks version .
Recall that the monogamy  of entanglement is an inequality that hints entanglements satisfy special constraints on how they distribute among multipartite systems.  The monogamy inequality  has  a nice
form: $$E_{A|BC}(\rho_{ABC})\geq E_{A|B}(\rho_{AB})+E_{A|C}(\rho_{AC}), \eqno(31)$$  with some entanglement measure $E(\cdot)$ \cite{Coffman}.
 The monogamy of
entanglement in quantum networks is key in quantum networks since design of entanglement distribution between multiple parties is the main scientific
challenge for realization of quantum networks \cite{Kimble}.
 However, most entanglement measures do not satisfy the monogamy inequality for entangled states. An interesting question is to find entangled states that allow the monogamy inequality \cite{Osborne,Oliveira,Luo4,Bai,Guo}. Our results shows the monogamy inequality is saturated for
preparable Gaussian entangled states (mixed or pure). This reveals the mechanism of entanglement distribution between multiple parties in Gaussian quantum networks. A parallel conclusion in discrete-variable systems is that
monogamy inequality under the entanglement of formation becomes equality for preparable pure entangled state of an $n$-partite quantum networks \cite{Luo4}.
\vspace{0.1in}

Although  mutual information of preparable entangled Gaussian states equals zero, these states actually have quantum correlation (beyond entanglement). An natural question is whether the monogamy inequality holds true for Gaussian quantum correlations. The following results shows that the answer is negative for preparable entangled Gaussian states.

In order to answer the question, one key step is to select a suitable Gaussian quantum correlation measure.
It is  found out the Gaussian quantum correlation measure from \cite{Hou1} is a good candidate for accurate description of desired correlation for preparable entangled Gaussian states.
Recall that for any $(n_1+n_2)-$mode $2-$partite Gaussian state $\rho_{A_1A_2}$, the quantity
$M(\rho_{A_1A_2})$ is defined by $$M(\rho_{A_1A_2})=1-\frac{\det V_{\rho_{A_1A_2}}}{ \det V_{\rho_{A_1}}\det V_{\rho_{A_2}}},\eqno(32)$$ where $V_{\rho_{A_1A_2}}$, $V_{\rho_{A_1}}$ and $V_{\rho_{A_2}}$
 are respectively the covariance matrices of $\rho_{A_1A_2}$, $\rho_{A_1}$ and $\rho_{A_2}$. It has been shown that $M(\cdot)$ is invariant under any permutation of subsystems, has no ancilla problem, is nonincreasing under $k$-partite
local Gaussian channels and vanishes on $k$-partite product states \cite{Hou1}. Based on the Gaussian quantum correlation measure $M(\cdot)$, we drive necessary criteria for entangled Gaussian states to be preparable in Gaussian quantum networks.
\vspace{0.1in}

\textbf{Theorem 5.} {\it For any preparable entangled Gaussian state $\rho$, the following statements hold true

(i) If $\rho_{XYZ}\in{\mathcal {GQN}}_\triangle$, then$$
   M_{{X|YZ}}(\rho_{XYZ})-M_{X|Y}(\rho_{{XY}})
- M_{X|Z}(\rho_{{XZ}})\leq 0.$$
\hspace{0.1in}(ii) If $\rho_{{A_1A_2\cdots A_{n+1}}}\in{\mathcal {GQN}}_\star$, then
$$\begin{array}{ll}M_{{A_1|A_2\cdots A_{n+1}}}(\rho_{{A_1A_2\cdots A_{n+1}}})&-M_{{A_1|A_2}}(\rho_{{A_1A_2}})-\cdots \\ &-M_{{A_1|A_{n+1}}}(\rho_{{A_1A_{n+1}}})\leq 0.\end{array}$$
(iii) If $\rho_{{A_1A_2\cdots A_{n+1}}}\in{\mathcal {GQN}}_\S$,  then
$$\begin{array}{ll}M_{{A_1|A_2\cdots A_{n+1}}}(\rho_{{A_1A_2\cdots A_{n+1}}})&-M_{{A_1|A_2}}(\rho_{{A_1A_2}})-\cdots \\ &-M_{{A_1|A_{n+1}}}(\rho_{{A_1A_{n+1}}})\leq 0.\end{array}$$}





\vspace{0.1in}

{\bf Proof.}  We only prove the star-shaped network case and  the other two cases can be proved similarly.
Since $M(\cdot)$ is invariant under locally Gaussian unitary operations, we assume $\rho=\rho_1\otimes \rho_2\otimes \cdots \otimes \rho_n$. For convenient computation, we can rewrite the covariance matrix of $\rho$ as the following form
$$\left(\begin{array}{ccccccccc}
\Gamma_{11}& 0 &\cdots &0& \Gamma_{1n+1} & 0 &\cdots &0\\
0& \Gamma_{22} &0 &\cdots &\cdots& \Gamma_{2n+2}  &\cdots &0\\
\vdots & \vdots  & \vdots &\ddots& \vdots & \vdots &\vdots& \vdots\\
\Gamma_{1n+1}^\dag & 0 & \cdots &0 &\Gamma_{n+1 n+1}& \cdots &0&0\\
\vdots & \vdots & \vdots &\vdots &\vdots& \ddots &\vdots&\vdots\\
0& 0& \cdots &\Gamma_{n 2n}^\dag &\cdots &0 &0&\Gamma_{2n 2n}\end{array}\right),\eqno(33)$$
each $\Gamma_{ii}$ corresponds the covariance matrix of $\rm{tr}_{A_{i+1}}\rho_i (1\leq i\leq n)$, $\Gamma_{n+in+i}$
is the covariance matrix of $\rm{tr}_{A_1}\rho_i (1\leq i\leq n)$, every off-diagonal block $\Gamma_{in+i}$ encodes the intermodal correlations between systems $i$ and $n+i$.
A direct computation shows that $$\begin{array}{ll}
 M(\rho_{A_1|A_2A_3\cdots A_{n+1}})&\\
=1-\frac{\det\left(\begin{array}{cc}
\Gamma_{11} & \Gamma_{1n+1}\\
\Gamma^\dag_{11} & \Gamma_{1n+1}\end{array}\right)\cdots\det\left(\begin{array}{cc}
\Gamma_{nn} & \Gamma_{n2n}\\
\Gamma^\dag_{n2n} & \Gamma_{2n2n}\end{array}\right)}{\det\Gamma_{11}\det\Gamma_{22}\cdots\det\Gamma_{2n2n}},&\end{array}\eqno(34)$$
$$ M(\rho_{A_1|A_2})=1-\frac{\det\left(\begin{array}{cc}
\Gamma_{11} & \Gamma_{1n+1}\\
\Gamma^\dag_{1n+1} & \Gamma_{n+1n+1}\end{array}\right)}{\det\Gamma_{11}\det\Gamma_{n+1n+1}},\eqno (35)$$
$$ M(\rho_{A_1|A_3})=1-\frac{\det\left(\begin{array}{cc}
\Gamma_{22} & \Gamma_{2n+2}\\
\Gamma^\dag_{2n+2} & \Gamma_{n+2n+2}\end{array}\right)}{\det\Gamma_{22}\det\Gamma_{n+2n+2}},\eqno (36)$$
$$\vdots$$
$$ M(\rho_{A_1|A_{n+1}})=1-\frac{\det\left(\begin{array}{cc}
\Gamma_{nn} & \Gamma_{n2n}\\
\Gamma^\dag_{n2n} & \Gamma_{2n2n}\end{array}\right)}{\det\Gamma_{nn}\det\Gamma_{2n2n}}.\eqno (37)$$
Let $$a_1=\frac{\det\left(\begin{array}{ll}
\Gamma_{11} & \Gamma_{1n+1}\\
\Gamma^\dag_{1n+1} & \Gamma_{n+1n+1}\end{array}\right)}{\det\Gamma_{11}\det\Gamma_{n+1n+1}},$$
$$a_2=\frac{\det\left(\begin{array}{cc}
\Gamma_{22} & \Gamma_{2n+2}\\
\Gamma^\dag_{2n+2} & \Gamma_{n+2n+2}\end{array}\right)}{\det\Gamma_{22}\det\Gamma_{n+2n+2}}$$
$$\vdots $$
$$ a_n=\frac{\det\left(\begin{array}{cc}
\Gamma_{nn} & \Gamma_{n2n}\\
\Gamma^\dag_{n2n} & \Gamma_{2n2n}\end{array}\right)}{\det\Gamma_{nn}\det\Gamma_{2n2n}},\eqno(38)$$
then each $a_i\in[0,1], 1\leq i\leq n$. By (32), a direct computation shows $$\begin{array}{ll}
 M(\rho_{A_1|A_2A_3\cdots A_{n+1}})-M(\rho_{A_1|A_2})-\cdots-M(\rho_{A_1|A_{n+1}})&\\
=1-n+a_1+\cdots +a_n-a_1a_2\cdots a_n&\end{array}\eqno(39)$$
In the following, we will show $$M(\rho_{A_1|A_2A_3\cdots A_{n+1}})-M(\rho_{A_1|A_2})-\cdots-M(\rho_{A_1|A_{n+1}})\leq 0.\eqno(40)$$
In fact, $(40)$  can be followed from an independent proposition: $$1-n+a_1+\cdots +a_n-a_1a_2\cdots a_n\leq 0\eqno(41)$$ for $a_i\in[0,1], 1\leq i\leq n$, which will be proved by the mathematical induction.
Firstly, for $n=2$, it is evident that $$-1+a_1+a_2-a_1a_2=(1-a_2)(a_1-1)\leq 0.\eqno(42)$$
Secondly, we deduce that $(41)$ is true for $n=k+1$
by assuming $(41)$ is true for $n=k$.  Indeed,
$$\begin{array}{ll}
                &-k+a_1+a_2+\cdots +a_{k+1}-a_1a_2\cdots a_{k+1}\\
             =  &1-k+a_1+\cdots +a_{k-1}+a_ka_{k+1}-a_1a_2\cdots a_{k+1}\\
                & +a_k+a_{k+1}-a_ka_{k+1}-1.\end{array}\eqno(43)$$
By our inductive assumption, for numbers $a_1, \cdots, a_{k-1}, a_ka_{k+1}$, we have $$1-k+a_1+\cdots +a_{k-1}+a_ka_{k+1}-a_1a_2\cdots  a_{k+1}\leq 0.\eqno(44)$$
Note that $$a_k+a_{k+1}-a_ka_{k+1}-1=(1-a_k)(a_{k+1}-1)\leq 0. \eqno(45)$$
Combining (44) with  (45), we have $$-k+a_1+a_2+\cdots +a_{k+1}-a_1a_2\cdots a_{k+1}\leq 0,\eqno(46)$$
which concludes the proof.
\vspace{0.1in}


Theorem 5 shows that any preparable Gaussian state disobeys the monogamy condition under Gaussian quantum correlation
measure $ M(\cdot)$.  The property is also a useful tool
 for detecting entangled Gaussian states which can
not be prepared by Gaussian quantum networks.
By Theorem 5, we can show that fully symmetric pure $4$-mode and $3$-partite Gaussian states can not be prepared by  Gaussian chain-type networks.
Indeed, a  $4$-mode fully symmetric  Gaussian state  has the symplectic spectrum
$$\begin{array}{ll}
v_-^2&=(b-e_1)(b-e_2),\\
v^2_{+^{(4)}}&=(b+3e_1)(b+3e_2),\end{array}\eqno (47)$$ $v_-$ is $3$ degenerate eigenvalue.
By Purity of states,  one can obtain
$$\begin{array}{ll}
e_1=&\{b^2-1+\sqrt{(b^2-1)(4b^2-1)}\}/{3b},\\
e_2=&\{b^2-1-\sqrt{(b^2-1)(4b^2-1)}\}/{3b}.\end{array}\eqno (48)$$
A direct computation shows that $$\begin{array}{ll}
 M(\rho_{{B|AC}})- M(\rho_{{B|A}})- M(\rho_{{B|C}})&\\
=\frac{v^2_{+^{(2)}}}{9}(2b^8-10b^6+17b^4-9),\end{array}\eqno (49)$$ here $v^2_{+^{(2)}}=(b+e_1)(b+e_2)$.
Note that $$f(b)=2b^8-10b^6+17b^4-9\  (b\geq 1)\eqno (50)$$ is strictly increasing, hence
$$ M_{B|AC}-M_{B|A}-M_{B|C}>0\ (b>1). \eqno (51)$$
Because that fully symmetric pure Gaussian states can be regarded as the vacuum  states if $b=1$,
thus pure $4$-mode fully symmetric entangled Gaussian state can not be generated by Gaussian chain-type networks. Fully symmetric pure Gaussian states belong to the class of continuous-variable Greenberger-Horne-Zeilinger (GHZ)-type state \cite{Adesso5,Loock}, an interesting result in discrete-variable systems is that GHZ states can not be prepared under local unitary operation \cite{Kraft}.

\vspace{0.1in}
${Conclusions\  and \ discussions }.$ Gaussian quantum networks are fundamental in network information theory. Here senders and receivers are connected through diverse routes that extend across intermediate sender-receiver pairs that act as nodes. The quantum
network is Gaussian if the operations at the nodes and the final state shared by end-users are Gaussian.
Gaussian quantum networks holds the huge potential to unlock more remarkable functions compared
to individual quantum systems. Given the deterministic
 preparation of quantum state high-speed measurement
capabilities, compatibility with classical fiber-optic networks,
 and cost-effectiveness, continuous-variable
quantum information is well-suited to the construction of Gaussian quantum networks \cite{Chengl}. Hence  preparation of entangled states is one of key elements for realizing Gaussian quantum networks.
Our goal  is to explore necessity criteria for preparing entangled stated in Gaussian quantum networks.  The necessity criteria show the network structure imposes strong constraints on the set
of preparable entangled Gaussian states. The necessity criteria also can be used as the witness for detecting entangled Gaussian states which can not be prepared by Gaussian quantum networks.

Our paper represents a natural starting point to build the Gaussian network entanglement theory.
Our results raise some interesting questions. It would be of
great interest to provide some sufficient conditions or informatics characterization for the
preparability of entangled quantum states in Gaussian quantum networks.
It is also interesting to discuss features of preparable entangled states in more general network topology. 
These questions are important to discover Gaussian network mechanism for preprable
entangled states.

\vspace{0.1in}
{\it Acknowledgement.---}
The authors thank sincerely all referees for thorough study of our manuscript and their valuable
feedback which is very helpful to improve the accuracy of our presentation.
The authors thank professor Shunlong Luo for helpful comments during the preparation of this paper. We acknowledge that this research was supported by NSF of China (12271452), NSF of Xiamen (3502Z202373018), and NSF of Fujian (2023J01028).

\end{document}